\documentclass[fleqn,twoside]{article}
\usepackage{espcrc2}

\usepackage{graphicx}
\usepackage[figuresright]{rotating}

\newcommand{\AmS}{{\protect\the\textfont2
  A\kern-.1667em\lower.5ex\hbox{M}\kern-.125emS}}

\newcommand{\comment}[1]{}

\def \rightdownarrow
 {\kern.3em
 \rule[.5ex]{.15mm}{2ex}
 {\mbox{$\kern-0.1em{\longrightarrow}$}}
 }
\def\lessim{\mathrel {\vcenter {\baselineskip 0pt \kern 0pt
\hbox{$<$} \kern 0pt \hbox{$\sim$} }}}
\def\gessim{\mathrel {\vcenter {\baselineskip 0pt \kern 0pt
\hbox{$>$} \kern 0pt \hbox{$\sim$} }}}
\newcommand{\beq}{\begin{equation}}
\newcommand{\eeq}{\end{equation}}
\newcommand{\bear}{\begin{array}}
\newcommand{\ear}{\end{array}}
\newcommand{\bet}{\begin{tabular}}
\newcommand{\eet}{\end{tabular}}
\newcommand{\beqn}{\begin{eqnarray}}
\newcommand{\eeqn}{\end{eqnarray}}

\newcommand{\bfh}{\begin{figure}[h]}
\newcommand{\efh}{\end{figure}[h]}

\newcommand{\tev}{\ensuremath{\mathrm{Te\kern -0.1em V}}}
\newcommand{\gev}{\ensuremath{\mathrm{Ge\kern -0.1em V}}}	
\newcommand{\mev}{\ensuremath{\mathrm{Me\kern -0.1em V}}}	
\newcommand{\kev}{\ensuremath{\mathrm{ke\kern -0.1em V}}}	
\newcommand{\massmev}{\mbox{\mev/$c^2$}}			
\newcommand{\pgev}{\mbox{\gev/$c$}}				

\newcommand{\CP}{CP}						
					
\newcommand{\pt}{\ensuremath{p_{\rm{T}}}}			
\newcommand{\ptb}{\ensuremath{\pt(B)}}				

\newcommand{\bd}{\ensuremath{B^{0}}}				
\newcommand{\bs}{\ensuremath{B^{0}_s}}				

\newcommand{\abd}{\ensuremath{\overline{B}^{0}}}		
\newcommand{\abs}{\ensuremath{\overline{B}^{0}_s}}		

\newcommand{\bn}{\ensuremath{B^{0}_{(s)}}}			
\newcommand{\bnmeson}{\mbox{$B^0_{(s)}$ meson}}			

\newcommand{\bhh}{\ensuremath{\bn \to h^{+}h^{'-}}}

\newcommand{\bshh}{\ensuremath{\bs \to h^{+}h^{'-}}}
\newcommand{\bdpipi}{\ensuremath{\bd \to \pi^+ \pi^-}}
\newcommand{\bdkpi}{\ensuremath{\bd \to K^+ \pi^-}}

\newcommand{\bskpi}{\ensuremath{\bs \to K^- \pi^+}}

\newcommand{\bskk}{\ensuremath{\bs \to  K^+ K^-}}
\newcommand{\bspipi}{\ensuremath{\bs \to  \pi^+ \pi^-}}
\newcommand{\bdkk}{\ensuremath{\bd \to  K^+ K^-}}

\newcommand{\etal}{{\em et al.}}

\newcommand{\Bd}{$B^{0}$}

\newcommand{\Bs}{$B_{s}^{0}$}
\newcommand{\Lb}{$\Lambda_{b}^{0}$}

\newcommand{\Bhh}{\ensuremath{\bn \to h^{+}h^{'-}}}
\newcommand{\Lbph}{\ensuremath{\Lambda^{0}_{b} \to ph^{-}}}
\newcommand{\Bdpipi}{\ensuremath{\bd \to \pi^+ \pi^-}}

\newcommand{\BdKpi}{\ensuremath{\bd \to K^+ \pi^-}}
\newcommand{\aBdKpi}{\ensuremath{\abd \to K^- \pi^+}}
\newcommand{\BsKpi}{\ensuremath{\bs \to K^- \pi^+}}
\newcommand{\aBsKpi}{\ensuremath{\abs\to K^+ \pi^-}}
\newcommand{\BsKK}{\ensuremath{\bs \to  K^+ K^-}}

\newcommand{\Bspipi}{\ensuremath{\bs \to  \pi^+ \pi^-}}

\newcommand{\BdKK}{\ensuremath{\bd \to  K^+ K^-}}

\newcommand{\Lbppi}{\ensuremath{\Lambda_{b}^{0} \to p\pi^{-}}}

\newcommand{\LbpK}{\ensuremath{\Lambda_{b}^{0} \to pK^{-}}}

\newcommand{\Dkpi}{\ensuremath{D^{0} \to K^- \pi^+}}

\newcommand{\DKpi}{\ensuremath{D^{0} \to K^- \pi^+}}
\newcommand{\aDKpi}{\ensuremath{\overline{D}^{0} \to K^+ \pi^-}}

\newcommand{\dedx}{\ensuremath{\rm{dE/dx}}}

\newcommand{\acpbdkpi}{\ensuremath{A_{\rm{\CP}}(\bdkpi)}}
\newcommand{\acpbukpi}{\ensuremath{A_{\rm{\CP}}(B^+ \to K^+ \pi^0)}}
\newcommand{\acpbskpi}{\ensuremath{A_{\rm{\CP}}(\bskpi)}}

\newcommand{\acpDKpi}{\ensuremath{A_{\rm{\CP}}(\DKpi)}}

\newcommand{\LbppisuLbpK}{\ensuremath{\mathcal{B}(\Lbppi)/\mathcal{B}(\LbpK)}}

\newcommand{\BR}{\ensuremath{\mathcal B}}
\newcommand{\cdf}{CDF Collaboration}

\newcommand{\ACPddef}{\ensuremath{{\frac{\BR (\aBdKpi)-\BR 
(\BdKpi)}{\BR (\aBdKpi)+\BR (\BdKpi)}}}}
\newcommand{\ACPsdef}{\ensuremath{{\frac{\BR (\aBsKpi)-\BR 
(\BsKpi)}{\BR (\aBsKpi)+\BR (\BsKpi)}}}}
\newcommand{\rateratiodef}{\ensuremath{\frac{\mathit{f_d}}{\mathit{f_s}}{\frac{ \Gamma(\aBdKpi)- 
\Gamma(\BdKpi)}{\Gamma(\aBsKpi)-\Gamma(\BsKpi)}}}}

\newcommand{\BdpipisuBdKpidef}{\ensuremath{\frac{\BR(\Bdpipi)}{\BR(\BdKpi)}}}
\newcommand{\BsKKsuBdKpidef}{\ensuremath{\frac{\mathit{f_s}}{\mathit{f_d}}\frac{\BR(\BsKK)}{\BR(\BdKpi)}}}

\newcommand{\BsKpisuBdKpidef}{\ensuremath{\frac{\mathit{f_s}}{\mathit{f_d}}\frac{\BR(\BsKpi)}{\BR(\BdKpi)}}}
\newcommand{\BspipisuBdKpidef}{\ensuremath{\frac{\mathit{f_s}}{\mathit{f_d}}\frac{\BR(\Bspipi)}{\BR(\BdKpi)}}}
\newcommand{\BdKKsuBdKpidef}{\ensuremath{\frac{\BR(\BdKK)}{\BR(\BdKpi)}}}

\newcommand{\LbppisuLbpKdef}{\ensuremath{\frac{\BR(\Lbppi)}{\BR(\LbpK)}}}

\newcommand{\Lumi}{\ensuremath{\mathcal{L}}}			
\newcommand{\lumifb}{\mbox{fb$^{-1}$}}				

\title{
Branching fractions and direct \CP\ asymmetries of charmless $B$ decay modes at the Tevatron
}
\author{M.~Morello\address[pisa]{Scuola Normale Superiore and  I.N.F.N di Pisa - Ed. C,
         Polo Fibonacci, Largo B. Pontecorvo, 3 - 56127 Pisa, Italy} \\
		 On behalf of the CDF Collaboration }

\begin{document}

\begin{abstract}
We present new CDF results on the branching fractions and time-integrated direct \CP\ 
asymmetries for \Bd\ and \Bs\ decay modes into pairs of charmless charged hadrons (pions or kaons). 
The data-set for this update amounts to 1~fb$^{-1}$ of $\bar{p}p$ collisions at a center of mass energy $1.96~\rm{TeV}$.
We report the first observation of the \BsKpi\ mode and a measurement
of its branching fraction and direct \CP\ asymmetry. We also observe for the first time two charmless decays 
of the $\Lambda_b$-baryon: \Lbppi\ and \LbpK.  
\vspace{1pc}
\end{abstract}

\maketitle

\section{INTRODUCTION}
The decay modes of  $B$ mesons into pairs of charmless
pseudo-scalar mesons are effective probes of the quark-mixing (CKM) matrix
and sensitive to potential new physics effects.
The large production cross section of $B$ hadrons of all kinds at the Tevatron
allows  measuring such decays in new modes, which
are important to supplement our understanding of $B$ meson decays.
The still unobserved \BsKpi\ mode could be used to measure
$\gamma$~\cite{Gronau:2000md} and its
\CP\ asymmetry could be a powerful model-independent test 
of the source of  direct
\CP\ asymmetry in the $B$ system \cite{Lipkin-BsKpi}. This may provide 
useful information to solve the
current discrepancy between the asymmetries observed in the neutral
\acpbdkpi\ and charged mode \acpbukpi~\cite{HFAG06}.

The \Bspipi\ and \BdKK\  modes proceed through annihilation and exchange topologies, which
are currently poorly known and a source of significant uncertainty in
many theoretical calculations~\cite{B-N,Bspipi}. 
A measurement of both modes would allow a determination of
the strength of these diagrams~\cite{Burasetal}.

CDF\,II is a multipurpose magnetic spectrometer surrounded by
calorimeters and muon detectors~\cite{CDF}. 
A silicon micro-strip detector (SVXII) and a cylindrical drift chamber
(COT) situated in a 1.4 T solenoidal magnetic field
reconstruct charged particles in the pseudo-rapidity range
$|\eta| < 1.0$.
The SVXII consists of five concentric layers
of double-sided silicon detectors with radii between 2.5 and 10.6 cm,
each providing a measurement with 15~$\mu$m resolution in the
azimuthal ($\phi$) direction and 70~$\mu$m along the beam ($z$) direction.
The COT has 96 measurement layers, between 40 and 137 cm in radius, 
organized into alternating axial and $\pm 2^{\circ}$ stereo ``super-layers''.
The transverse momentum resolution is $\sigma_{p_{T}}/p_{T} \simeq
0.15\%\, p_{T}$/(GeV/$c$) and the observed mass-widths are about 14 \massmev\
for $J/\psi\to\mu^+\mu^-$ decays, and about 9 \massmev\ for \Dkpi\ decays.
 The specific energy loss by ionization (\dedx) of charged particles in the COT can
be measured from the amount of charge collected by each wire.

Throughout this paper, C-conjugate modes
are implied and branching fractions indicate
\CP-averages unless otherwise stated.

\section{\label{sec:Data}DATA SAMPLE}
We analysed an integrated luminosity  $\int\Lumi dt\simeq 1$~\lumifb\ sample of pairs of oppositely-charged particles
with $p_{T} > 2$~\pgev\ and   $p_{T}(1) + p_{T}(2) > 5.5$~\pgev,
used to form \bnmeson\ candidates.
The trigger required also a transverse opening angle $20^\circ < \Delta\phi < 135^\circ$ between the two tracks,
  to reject background from particle 
pairs within the same jet and from back-to-back jets.
In addition, both charged particles were required to originate from
a displaced vertex with a large impact parameter $d_0$ (100 $\mu$m $< d_0 < 1$~mm), 
while the \bnmeson\ candidate was required to be produced in
the primary $\bar{p}p$ interaction ($d_0(B)< 140$~$\mu$m) and to have travelled a transverse distance
$L_{xy}(B)>200$~$\mu$m.

In the offline analysis, an unbiased optimization procedure determined a
 tightened selection on track-pairs fit to a common decay vertex.
We chose the selection cuts minimizing directly the expected uncertainty of the physics
observables to be measured (through 
several ``pseudo-experiments'').
We used just two different sets of cuts, respectively optimized
to measure the \CP\ asymmetry \acpbdkpi\ and the branching fraction \BR(\BsKpi),
since those two measurements are the main focus of the analysis. 
For the latter, the sensitivity for discovery and limit
setting~\cite{gp0308063} was
optimized rather than the statistical uncertainty on the particular parameter, since
the mode had not yet been observed.
It was verified that the former set of cuts is also adequate to measure 
other decay rates of the larger yield modes
(\bdpipi, \bskk), while the latter, tighter
set of cuts, is well suited to measure the decay rates and \CP\ asymmetries related to rare modes
(\bspipi, \bdkk, \Lbppi\ and \LbpK).

\begin{figure}[htb]
\includegraphics[scale=0.35]{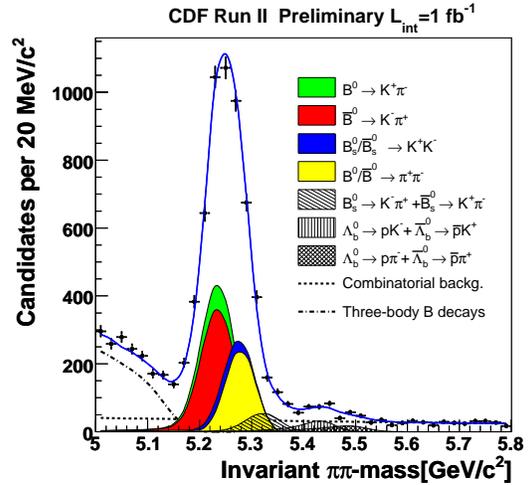}
\caption{Invariant mass distribution
of \bhh\ candidates passing all selection requirements optimized to measure \BR(\BsKpi), 
using a pion mass assumption for both decay products.
Cumulative projections of the likelihood fit for each mode are
overlaid.}
\label{fig:projections}
\end{figure}

In addition to tightening the trigger cuts, in the offline analysis the discriminating power
of the \bnmeson\ isolation and of the information provided by the 3D reconstruction 
capability  of the CDF tracking were used,
allowing a great improvement in the signal purity.
 Isolation is defined as $I(B)= \ptb/[\ptb + \sum_{i} \pt(i)]$, in which the sum runs over every other track 
(not from the $B$ meson) within
a cone of unit radius in the $\eta-\phi$ space around the \bnmeson\ flight direction. 
By requiring $I(B)> 0.5$,
  we reduced the background by a factor
four while keeping almost 80\% of signal. The 3D silicon tracking allowed  multiple vertices  to be resolved
along the beam direction and the rejection of fake tracks, reducing the background
by a factor of two, with only a small efficiency loss on signal.
The resulting $\pi\pi$-mass distributions (see Fig.~\ref{fig:projections}) show a clean signal of \bhh\ decays.
In spite of a good mass resolution ($\approx 22\,\massmev$), the various \bhh\ modes overlap into an unresolved
mass peak.

\section{\label{sec:fit}FIT TO THE $B$ DECAY-MODE COMPOSITION}
\begin{figure}[htb]
\begin{center}
\includegraphics[scale=0.25]{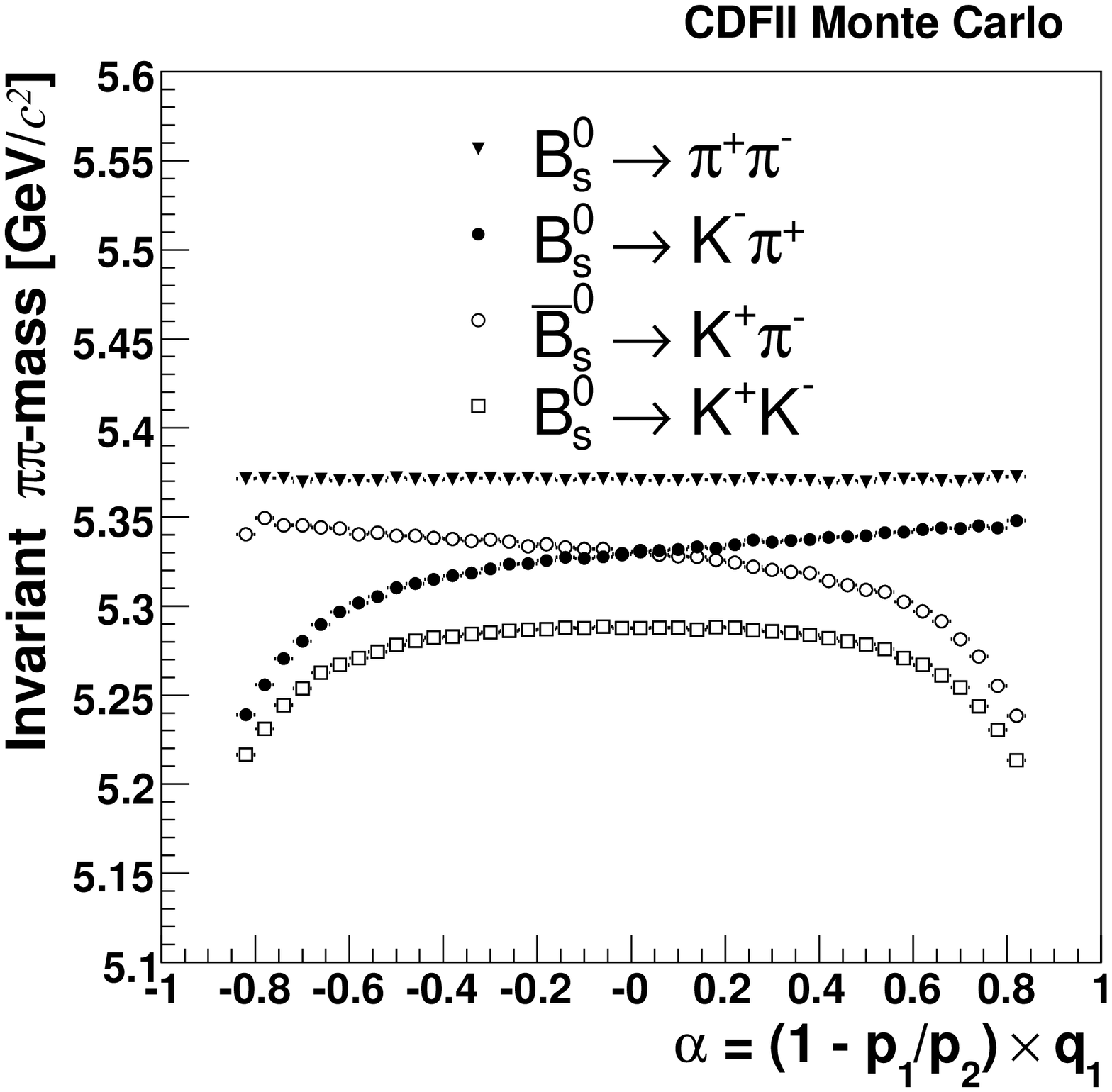}
\includegraphics[scale=0.25]{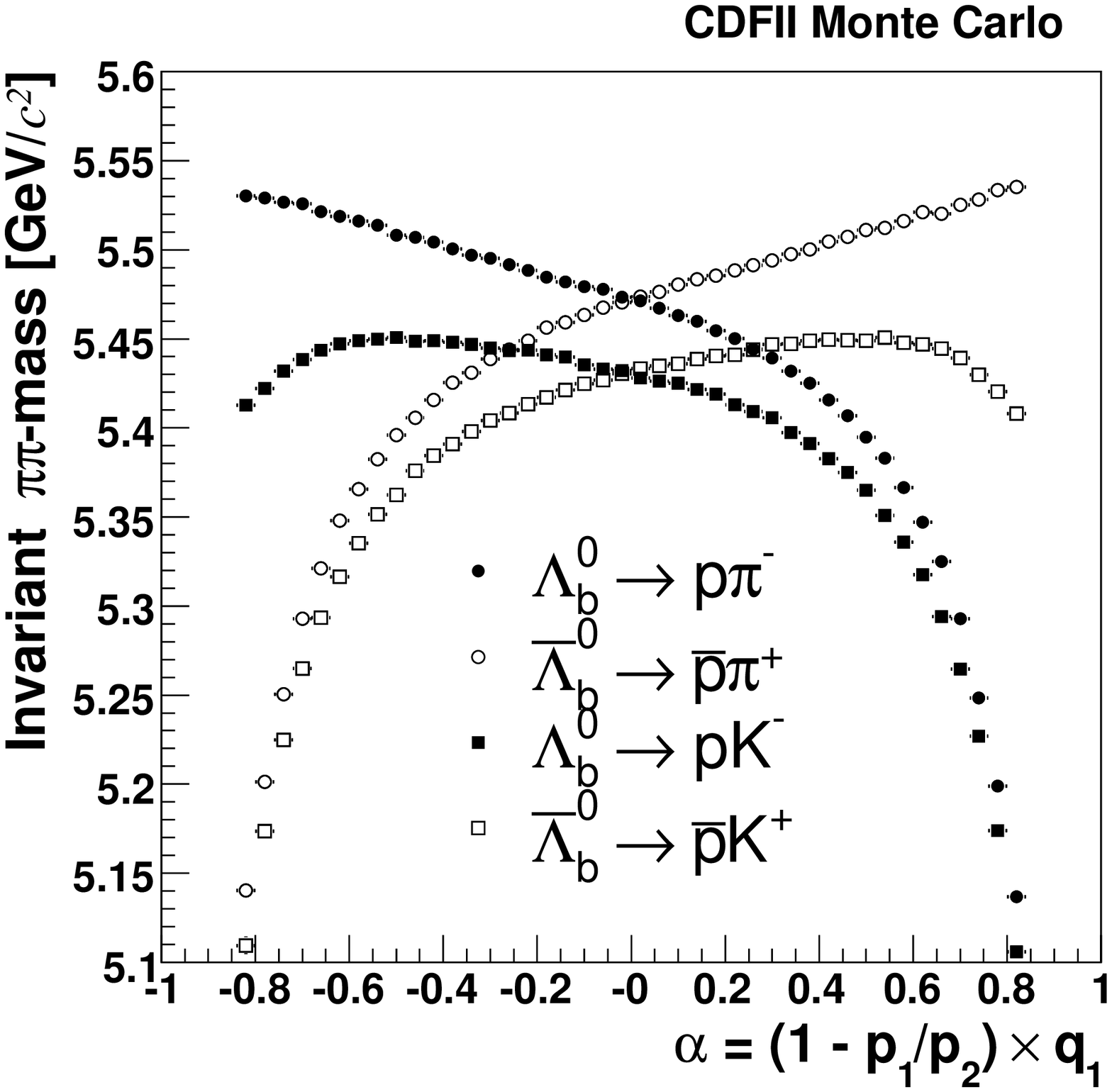}
\caption{Average $M_{\pi\pi}$ versus $\alpha$ for simulated
samples of \bs\ (top) and \Lb\ (bottom) candidates.
The corresponding plots for the \bd\ are similar to \bs\, but shifted for the mass difference.}
\label{fig:mpipi_vs_alpha}
\end{center}
\end{figure}

The resolution in invariant mass and in particle identification is not 
sufficient for separating the individual decay modes on an event-by-event basis,
therefore we performed an unbinned maximum likelihood fit, combining kinematic and particle identification information
to statistically determine both the contribution of each mode,
and the relative contributions to the \CP\ asymmetries.
For the kinematic portion, we used three loosely correlated
observables to summarize the
information carried by all possible values of invariant mass of the
$B$ candidate,
resulting from different mass assignments to the two outgoing
particles~\cite{punzibias}. They are: (a) the mass
$M_{\pi\pi}$ calculated with the charged pion mass assignment to both
particles; (b) the signed momentum imbalance
$\alpha = (1-p_1/p_2) q_{1}$, where $p_1$ ($p_2$) is the
lower (higher) of the particle momenta, and $q_1$ is the sign of the charge of the
particle of momentum $p_{1}$; (c) the scalar sum of the particle momenta $p_{tot}=p_1 + p_2$.
Using these three variables, the mass of any particular mode $M_{12}$
can be written  as:
\begin{eqnarray}\label{eq:Mpipi2}
M^{2}_{12}  &= & M^{2}_{\pi\pi}  -  2 m_{\pi}^2 +(m_{1}^2+m_{2}^2)                                 \nonumber \\
            &  &                 -  2 \sqrt{p_{1}^2+m_{\pi}^2} \cdot \sqrt{p_{2}^2+m_{\pi}^2}      \nonumber \\
            &  &                 -  2\sqrt{p_{1}^2+m_{1}^2} \cdot \sqrt{p_{2}^2+m_{2}^2},
\end{eqnarray}
\vspace{-0.5cm}
\begin{equation}\label{eq:sostituzione}
p_1 = \frac{1-|\alpha|}{2-|\alpha|}p_{tot} ~,~p_2 = \frac{1}{2-|\alpha|}p_{tot},
\end{equation}
where $m_{1}$ ($m_{2}$) is the mass of the lower (higher) momentum
particle. For simplicity, Eq.~(\ref{eq:Mpipi2}) is written
as a function of $p_{1}$ and $p_{2}$,
but in the likelihood it was used as a function of $\alpha$ and $p_{tot}$. 
The simulated average values of $M_{\pi\pi}$ as a function of $\alpha$ for the eight 
\bshh\ and \Lbph\ modes are shown in 
 Fig.~\ref{fig:mpipi_vs_alpha}.

Particle identification (PID) information  is
summarized by a single observable $\kappa_{1(2)}$ for track $1(2)$, 
defined as 
$\frac{dE/dx_{1(2)} - dE/dx_{1(2)}(\pi)}{dE/dx_{1(2)}(K) - dE/dx_{1(2)}(\pi)}$,
where $dE/dx_{1(2)}(\pi)$ and $dE/dx_{1(2)}(K)$ are the expected $dE/dx_{1(2)}$ depositions for those particle assignments.
With the chosen observables, the likelihood
 contribution of the $i^{\mathit{th}}$ event is written as:
 \begin{eqnarray}\label{eq:likelihood}
    \mathcal{L}_i & = & (1-b)\sum_{j} f_j \mathcal{L}^{\mathrm{kin}}_j  \mathcal{L}^{\mathrm{PID}}_j \nonumber \\
                  &   & +  b \left( f_{\rm{A}} \mathcal{L}^{\mathrm{kin}}_{\mathrm{A}}
    \mathcal{L}^{\mathrm{PID}}_{\mathrm{A}}+
   (1-f_{\rm{A}}) \mathcal{L}^{\mathrm{kin}}_{\mathrm{E}}
    \mathcal{L}^{\mathrm{PID}}_{\mathrm{E}}
	\right)
\end{eqnarray}
where:
\begin{equation}
    \label{eq:signal}\mathcal{L}_j^{\mathrm{kin}}=
     R(M_{\pi\pi}|\alpha,p_{tot})
     P_{j}(\alpha,p_{\rm{tot}}),
\end{equation}
 \begin{equation}\label{eq:bck_A}
\mathcal{L}^{\mathrm{kin}}_{\mathrm{A}} =
           {\rm A}(M_{\pi\pi};c_{2},m_{0}) P_{\mathrm{A}}(\alpha,p_{\rm{tot}}),
\end{equation}
\begin{equation}
\label{eq:bck_E}\mathcal{L}^{\mathrm{kin}}_{\mathrm{E}} =
           e^{c_{1} M_{\pi\pi}} P_{\mathrm{E}}(\alpha,p_{\rm{tot}}),
\end{equation}
\begin{equation}
\label{eq:PID_sig} \mathcal{L}^{\mathrm{PID}}_{j} =
   	F_{j}(\kappa_{1}, \kappa_{2}|\alpha,p_{\rm{tot}}),
\end{equation}
\begin{equation}
\label{eq:PID_bg} \mathcal{L}^{\mathrm{PID}}_{\mathrm{A(E)}} =
        \sum_{l,m=e,\pi,K,p} w^{\mathrm{A(E)}}_{l} w^{\mathrm{A(E)}}_{m} F_{lm}(\kappa_{1}, \kappa_{2}|\alpha,p_{\rm{tot}}).
\end{equation}
The various terms of the likelihood functions are described below.

The index $j$ runs over the twelve distinguishable \Bhh\ and \Lbph\ modes, and
$f_j$ are their fractions to be determined by the fit, together with the total background fraction $b$.
The background is composed of two different kinds: combinatorial background and 
partially-reconstructed heavy flavor decays. The combinatorial background is composed of random pairs 
of charged particle, displaced from the beam-line, accidentally satisfying the selection requirements,
while the  latter, referred as ``physics'' background, is composed of multi-body $b$-hadron decays
(i.e.~$\bn \to \rho\pi/\rho K$) in which only two tracks are reconstructed.   
The indices $\mathrm{A(E)}$ label the physics (combinatorial) background quantities.
The fraction of the physics background is given by $f_{A}$ and it is a free parameter in the fit.

Each likelihood term, both for signals and backgrounds, is factorized into three different contributions:
a) the conditional probability distribution of the invariant mass $M_{\pi\pi}$ given
$\alpha$ and $p_{tot}$ (for the background $M_{\pi\pi}$ is assumed to be independent of momentum), 
b) the joint conditional probability of PID variables 
$\mathcal{\kappa}_1$, $\mathcal{\kappa}_2$ given $\alpha$, $p_{tot}$ for a determined particles hypothesis, $j$ in the case of 
signals ($F_{j}$) and $l,m$ in the case of background ($F_{l,m}$), and 
c) the joint probability distribution of momentum variables $\alpha$ and $p_{tot}$ ($P_{j\mathrm{(A,E)}}$).

$R(M_{\pi\pi}|\alpha,p_{tot}) = R(M_{\pi\pi}-\mathcal{M}_{j}(\alpha,p_{tot}),\alpha,p_{tot})$
is the mass resolution function of each mode $j$ when the correct mass is assigned to both tracks. 
The average mass $\mathcal{M}_{j}(\alpha,p_{tot})$ is the value of $M_{\pi\pi}$ obtained 
from Eq.~(\ref{eq:Mpipi2}) by setting the appropriate particle masses for each decay 
mode $j$ and, by  making a simple variable change, we obtain
$R(M_{\pi\pi}-\mathcal{M}_{j}(\alpha,p_{tot}),\alpha,p_{tot})= R(M_{j}-M_{B^0(B^{0}_{s},\Lambda^0_b)},\alpha,p_{tot})$,
where $M_{j}$ is the  invariant mass computed with the correct mass assignment to both particles for each mode $j$.
 $\alpha$ and $p_{tot}$ appear explicitly in the last equation since they are useful to parameterize the dependence of 
 the mass resolution by the momenta.

The mass distribution of the physics background is parameterized with  
an ``Argus function'', defined by the notation
${\rm A}(M_{\pi\pi};c_{2},m_{0})$~\cite{argus}, convoluted with a Gaussian distribution
centered at zero with a width, in this case, equal to the mass resolution,
while the combinatorial background with an exponential function. 
The background mass distribution was determined in
the fit by varying the parameters $c_1$, $c_2$ and $m_{0}$ in Eq.~(\ref{eq:bck_A},\ref{eq:bck_E}).
The function $P_{j\rm{(A,E)}}(\alpha,p_{tot})$ 
was parameterized by a product of
polynomial and exponential functions fitted to Monte Carlo samples
produced by a detailed detector simulation for each mode $j$, instead for the background 
 terms was obtained from the mass sidebands of data.
 
\begin{figure}[tb]
\includegraphics[scale=0.3]{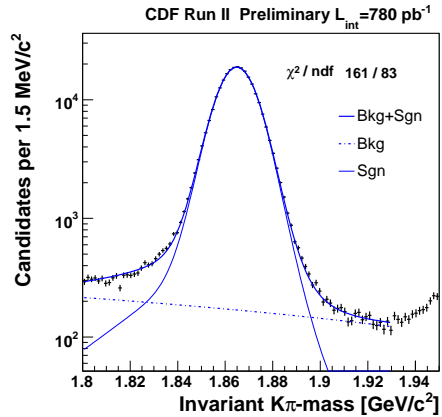}
\caption{Tagged \Dkpi\  decays from $\ensuremath{D^{*+} \to D^{0} \pi^{+} \to [K^{-}\pi^{+}]\pi^{+}}$: 
a verification of the mass line shape 
by performing a 1-D binned fit where the
signal mass line shape is completely fixed from the model (see text). 
}
\label{fig:dstar1}
\end{figure}

The mass resolution function $R$ was parameterized using the detailed detector simulation.
To take into account non-Gaussian tails due to the emission of photons in the final state, 
we included  soft photon emission in the simulation, using recent QED calculations \cite{Cirigliano-Isidori}.
The quality of the mass resolution model was verified  in the simulation
using about 500k 
\DKpi\ decays (see Fig.~\ref{fig:dstar1}). 
\begin{figure}[tb]
\includegraphics[scale=0.3]{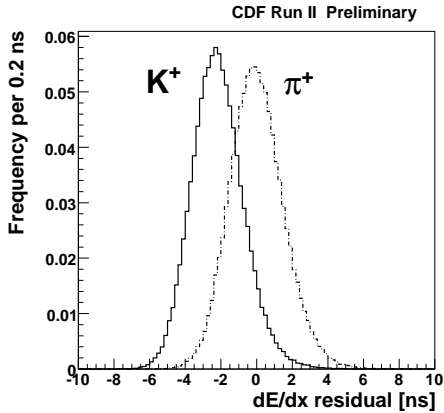}
\caption{Tagged \Dkpi\  decays from $\ensuremath{D^{*+} \to D^{0} \pi^{+} \to [K^{-}\pi^{+}]\pi^{+}}$: 
 distribution of
\dedx\ around the average pion response,
for calibration samples of kaons and pions. COT samples the amount of ionization charge
produced by a track by measuring the time-over-threshold (ns) of the pulse on each wire associated 
to the track.
}
\label{fig:dstar2}
\end{figure}
The mass line-shape of the \DKpi\ 
was fitted by fixing the signal 
shape from the model, and allowing to vary only the background function.
Good agreement was obtained between data and simulation.
In Eq.~(\ref{eq:signal}),  the nominal \Bd, \Bs\ and \Lb\ masses measured by CDF~\cite{CDFmasspaper} were used
to reduce the systematic uncertainties related to the knowledge of the global mass scale.

A sample of 1.5M $D^{*+}\to D^0\pi^+
 \to [K^-\pi^+]\pi^+$ decays, where the $D^0$ decay
products are identified by the charge of the $D^{*+}$ pion,
was used to calibrate the \dedx\ response over the tracking volume
and over time, and to determine the $F_{j(l,m)}(\mathcal{\kappa}_1,\mathcal{\kappa}_2|\alpha,p_{tot})$ functions
in Eq.~(\ref{eq:PID_sig},\ref{eq:PID_bg}). 
In a $>95\%$ pure $D^0$ sample, we obtained $1.4\sigma$ separation between kaons and pions
(see Fig.~\ref{fig:dstar2}).
The PID background term in Eq.~(\ref{eq:PID_bg}) 
is similar to the signal terms, but allows for independent pion, kaon, proton, and electron components, 
which are free to vary independently for physics (combinatorial) background. 
In Eq.~(\ref{eq:PID_bg}) the indices $l$ and $m$ run over the four possible particles 
$e,~\pi,~K,~p$ and the fractions of different kind of particles $w^{\rm{A(E)}}_{l}$,$w^{\rm{A(E)}}_{m}$
are free parameters in the fit. Muons are indistinguishable from pions with the
available \dedx\ resolution.

\section{\label{sec:fitresults} FIT RESULTS AND SYSTEMATICS}

We performed two separate fits: the first  using the cuts
optimized to measure the direct \acpbdkpi\ and
the second  to measure \BR(\BsKpi).
Significant signals are seen for \Bdpipi, \BdKpi, and \BsKK, previously observed by CDF~\cite{paper_bhh}.
Three new rare modes were observed for the first time: \BsKpi, \Lbppi\ and \LbpK\
while no evidence was obtained for \Bspipi or  \BdKK.

\begin{table*}
\caption{\label{tab:summary}Results on data sample optimized to measure \acpbdkpi\ (top) and
 \BR(\BsKpi) (bottom). Absolute branching fractions are normalized to the the world--average values
${\mathcal B}(\mbox{\BdKpi}) = (19.7\pm 0.6) \times 10^{-6}$ and
$f_{s}= (10.4 \pm 1.4)\%$ and $ f_{d}= (39.8 \pm 1.0)\%$~\cite{HFAG06}.
The first quoted uncertainty is statistical, the second is systematic.
$N_s$ is the number of fitted events for each mode. For rare modes both systematic and statistical uncertainty on $N_s$ was quoted 
while for abundant modes only statistical one. For the \Lb\ modes only the ratio \LbppisuLbpKdef\ was measured.}
{\footnotesize
\begin{tabular}{lc|lc|c}
\hline
Mode & N$_{s}$ & Quantity & Measurement & \BR (10$^{-6}$)  \\
\hline
\BdKpi         & 4045 $\pm$ 84          &  \ACPddef\         & -0.086 $\pm$ 0.023 $\pm$ 0.009   &                            \\
\Bdpipi        & 1121 $\pm$ 63          & \BdpipisuBdKpidef\ & 0.259 $\pm$ 0.017 $\pm$ 0.016    & 5.10 $\pm$ 0.33 $\pm$ 0.36 \\
\BsKK          & 1307 $\pm$ 64          & \BsKKsuBdKpidef\   &  0.324 $\pm$ 0.019 $\pm$ 0.041   & 24.4 $\pm$ 1.4 $\pm$ 4.6   \\
\hline
\BsKpi         & 230 $\pm$ 34 $\pm$ 16  & \BsKpisuBdKpidef\  &  0.066 $\pm$ 0.010 $\pm$ 0.010   & 5.0 $\pm$ 0.75 $\pm$ 1.0   \\
               &                        &  \ACPsdef\         &  0.39 $\pm$ 0.15 $\pm$ 0.08      &                            \\
               &                        &  \rateratiodef\    &  -3.21 $\pm$ 1.60 $\pm$ 0.39     &                            \\
\Bspipi        & 26 $\pm$ 16 $\pm$ 14   &\BspipisuBdKpidef\  &  0.007 $\pm$ 0.004 $\pm$ 0.005   & 0.53 $\pm$ 0.31 $\pm$ 0.40 \\
	       &                        &      		     &  		                & ($< 1.36$ @~90\%~CL)       \\
\BdKK          & 61 $\pm$ 25  $\pm$ 35  & \BdKKsuBdKpidef\   &  0.020 $\pm$ 0.008 $\pm$ 0.006   & 0.39 $\pm$ 0.16 $\pm$ 0.12 \\
	       &                        &      		     &  	                        &  ($< 0.7$ @~90\%~CL)       \\
\LbpK          & 156 $\pm$ 20 $\pm$ 11  &\LbppisuLbpKdef\    &  0.66 $\pm$ 0.14 $\pm$ 0.08      &                            \\
\Lbppi         & 110 $\pm$ 18 $\pm$ 16  &                    &                                  &                            \\
\hline
\end{tabular}
}
\end{table*}

To convert the yields returned from the fit into relative branching fractions,
we applied corrections for efficiencies of trigger and
offline selection requirements for different decay modes.
The relative efficiency corrections between modes do not exceed $20\%$.
Most corrections were determined from the detailed
detector simulation, with some exceptions which were measured using data.
A momentum-averaged relative isolation efficiency between \Bs\
and \Bd\ of $1.07 \pm 0.11$  was determined from fully-reconstructed samples of
\Bs $\to J/\psi\,\phi$, \Bs $\to D^{-}_{s}\pi^{+}$,
\Bd $\to J/\psi\,K^{*0}$,
and \Bd $\to D^{-}\pi^{+}$.
The lower specific ionization of kaons with respect to pions in the
drift chamber is responsible for a
$\simeq 5$\% lower efficiency to reconstruct a kaon.
This effect was measured in a sample of $D^{+}\to K^{-}\pi^{+}\pi^{+}$
decays triggered on two tracks, using the unbiased third track.
The only correction needed for the direct \CP\ asymmetries \acpbdkpi\ and
\acpbskpi\ was a $\le 0.6\%$ shift due to the different probability for $K^{+}$ and
$K^{-}$ to interact with the tracker material. The measurement of this correction
has been achieved using a sample of 1M of prompt \DKpi\ decays reconstructed and selected
using the same criteria as \Bhh\ decays. Assuming the Standard Model expectation of $\acpDKpi=0$, 
the difference between  the number of reconstructed \DKpi\ and \aDKpi\ 
 provides a measurement of the 
detector-induced asymmetry between $K^+\pi^-$ and $K^-\pi^+$ final states.
Since the same fit technique developed for the \Bhh\ decays was used,
this measurement provides also a robust check on all possible charge asymmetry biases of the detector
and \dedx\ parameterizations. 

The \BsKK\ and \Bspipi\ modes required a special treatment, since they
contain a superposition of the flavor eigenstates of the \Bs .
Their time evolution might differ from the one of the flavor-specific
modes if the width difference $\Delta\Gamma_{s}$
between the \Bs\ mass eigenstates is significant.
The current result was derived under the assumption that both modes are
dominated by the short-lived \Bs\ component, that $\Gamma_s=\Gamma_d$, and $\Delta\Gamma_s/\Gamma_s =
0.12\pm 0.06$~\cite{Beneke:1998sy,Lenz:2004nx}. 
 The latter uncertainty is included in estimating the overall systematic
uncertainty.

The dominant contributions to the systematic uncertainty are
the statistical uncertainty on the isolation efficiency (\Bs\ modes),
the uncertainty on the \dedx\ calibration and parameterization, and the uncertainty of the combinatorial
background model.
The first  is the larger systematic of all measurements with a  $B^0_s$ meson decay
(except for \acpbskpi). 
The second systematic, due to \dedx, is a large systematic of all measurements,
although the parameterization of the \dedx\ is very accurate.
The fit of composition is very sensitive to the PID information. 
The third one is due to the statistical uncertainty
of the possible combinatorial background models and it is a dominant systematic for the
observables of the rare modes.
Smaller systematic uncertainties are assigned for trigger efficiencies, physics background shape, kinematics,
$B$ meson masses and lifetimes.

\section{\label{sec:results}RESULTS}
The relative branching fractions 
are listed in Table~\ref{tab:summary}, where $f_{d}$ and $f_{s}$ indicate
the production fractions respectively of \Bd\ and \Bs\
from fragmentation of a $b$ quark in $\bar{p}p$ collisions.
An upper limit is also quoted for modes in which no significant signal is
observed~\cite{F-C}. We also list absolute results obtained by normalizing the data to 
the world-average of \BR(\BdKpi)~\cite{HFAG06}. The contributions
from the likelihood fit for each decay mode are shown in Fig.~\ref{fig:projections}.

We report the first observation of three new rare charmless decays \BsKpi, \Lbppi\ and \LbpK\
with a significance respectively of $8.2 \sigma$, $6.0 \sigma$ and $11.5 \sigma$. 
The significance includes both statistical and systematic uncertainty. The statistical uncertainty 
to evaluate the significance was estimated using several pseudo-experiments with no contributions 
from rare signals. 

The branching fraction of the newly observed mode 
$\BR(\BsKpi)=(5.0 \pm 0.75 \pm 1.0)\times 10^{-6}$ is in agreement with the latest
theoretical expectation \cite{zupan} which is lower than  the previous predictions \cite{B-N,Yu-Li-Cai}.
We measured for the first time in the \bs\ meson system
the direct \CP\ asymmetry of $\acpbskpi=0.39 \pm 0.15 \pm 0.08$.
This value favors a large \CP\ violation in  \bs\ meson decays, conversely
it is also compatible with zero. 
In Ref.~\cite{Lipkin-BsKpi}  a robust test of the Standard Model or a probe of new physics is suggested by
comparison of the direct \CP\ asymmetries in  \BsKpi\ and \BdKpi\ decays.
Using HFAG input~\cite{HFAG06} we measure  $\frac{\Gamma(\aBdKpi)-\Gamma(\BdKpi)}{\Gamma(\BsKpi)-\Gamma(\aBsKpi)}  
= 0.84 \pm 0.42 \pm 0.15$ where $\Gamma$ is the decay width, 
in agreement with the Standard Model expectation of unity. Assuming that the relationship above is unity 
and using as input the \BR(\BsKpi) measured here, the world average for \acpbdkpi\  and  \BR(\BdKpi)~\cite{HFAG06}, 
we can estimate the expected value for
$\acpbskpi \approx 0.37$ in agreement with our measurement. 

The branching fraction $\BR(\BsKK)= (24.4 \pm 1.4 \pm 4.6) \times 10^{-6}$ is in agreement with the 
latest theoretical expectation~\cite{matiasBsKK,matiasBsKK2}
and with the previous CDF measurement \cite{paper_bhh}.
An improved systematic uncertainty is 
expected for the final analysis of the same sample.

The results for the \Bd\ are in agreement with world average values~\cite{HFAG06}.
The measurement $\acpbdkpi =-0.086 \pm 0.023 \pm 0.009$ is the world's second best measurement
and the significance of the new world average $A^{ave.}_{\mathsf{CP}}(\bdkpi)=-0.095 \pm 0.013$
moved from 6$\sigma$ to 7$\sigma$.

The updated upper limits and the absolute branching fractions of the currently  unobserved 
modes \BdKK\ and \Bspipi\ have been reported.  
The rate $\BR(\BdKK) = (0.39 \pm 0.16 \pm 0.12)\times 10^{-6}$  has the same uncertainty of 
the current measurements~\cite{HFAG06} while
the \Bspipi\ upper limit (already the world's best~\cite{paper_bhh}) was improved by a factor 1.3,  
approaching the expectations from recent calculations~\cite{Bspipi,Bspipi-PQCD}.

We also report the first observation of  two new baryon charmless modes \Lbppi\ and \LbpK. 
We measured $\LbppisuLbpK =  0.66 \pm 0.14 \pm 0.08$, in agreement with the expectations from \cite{Mohanta}. 

\end{document}